\documentclass[
  reprint,
superscriptaddress,
 amsmath,amssymb,
 aps,
prb,
]{revtex4-2}

\usepackage{graphicx}
\usepackage{dcolumn}
\usepackage{bm}
\usepackage{blindtext}
\usepackage{float}
\usepackage{caption,cancel}
\usepackage{cleveref}
\usepackage{subcaption}
\usepackage{xcolor}
\usepackage{pdfpages}

\makeatletter
\AtBeginDocument{\let\LS@rot\@undefined}
\makeatother

\newcommand{\bom}{\boldsymbol{\omega}}

\def \s{\mathbf{s}}
\def \v{\mathbf{v}}
\def \x{\mathbf{x}}

\begin{document}


\title{Experimental and theoretical evidence of universality in superfluid vortex reconnections}

\author{P. Z. Stasiak}
\affiliation{School of Mathematics, Statistics and Physics, Newcastle University, Newcastle upon Tyne, NE1 7RU, United Kingdom}

\author{Y. Xing}
\author{Y. Alihosseini}
\affiliation{National High Magnetic Field Laboratory, 1800 East Paul Dirac Drive, Tallahassee, Florida 32310, USA}
\affiliation{Mechanical Engineering Department, FAMU-FSU College of Engineering, Tallahassee, Florida 32310, USA}

\author{C.F. Barenghi}
\affiliation{School of Mathematics, Statistics and Physics, Newcastle University, Newcastle upon Tyne, NE1 7RU, United Kingdom}

\author{A. Baggaley}
\affiliation{Department of Mathematics and Statistics, Lancaster University, Lancaster, LA1 4YF, United Kingdom}
\affiliation{School of Mathematics, Statistics and Physics, Newcastle University, Newcastle upon Tyne, NE1 7RU, United Kingdom}

\author{W. Guo}
\affiliation{National High Magnetic Field Laboratory, 1800 East Paul Dirac Drive, Tallahassee, Florida 32310, USA}
\affiliation{Mechanical Engineering Department, FAMU-FSU College of Engineering, Tallahassee, Florida 32310, USA}

\author{L. Galantucci}
\affiliation{Istituto per le Applicazioni del Calcolo ``M. Picone" IAC CNR, Via dei Taurini 19, 00185 Roma, Italy}

\author{G. Krstulovic}
\affiliation{Universit\'e C\^ote d'Azur, Observatoire de la C\^ote d'Azur, CNRS,Laboratoire Lagrangre, Boulevard de l'Observatoire CS 34229 - F 06304 NICE Cedex 4, France}


\begin{abstract}
The minimum separation between reconnecting vortices
in fluids and superfluids obeys a universal scaling law with respect to time.
The pre-reconnection and the post-reconnection prefactors 
of this scaling law are different, a property related to irreversibility and to energy
transfer and dissipation mechanisms.
In the present work, we determine the temperature dependence of these prefactors in superfluid helium
from experiments and a numeric model which fully accounts for the 
coupled dynamics of the superfluid
vortex lines and the thermal normal fluid component. At all temperatures, we observe 
a pre- and post-reconnection asymmetry similar to that observed in other superfluids
and in classical viscous fluids, indicating that vortex reconnections display a
universal behaviour independent of 
the small-scale regularising dynamics.   
We also numerically show that each vortex reconnection event
represents a sudden injection of energy in the normal fluid. 
Finally we argue that
in a turbulent flow, these punctuated energy injections can sustain 
the normal fluid
in a perturbed state, provided that the density of superfluid vortices is large enough.
\end{abstract}

\maketitle

\begin{figure*}[t]
	\centering
	\begin{subfigure}[b]{0.24\textwidth}
		\centering
		\includegraphics*[width=\textwidth]{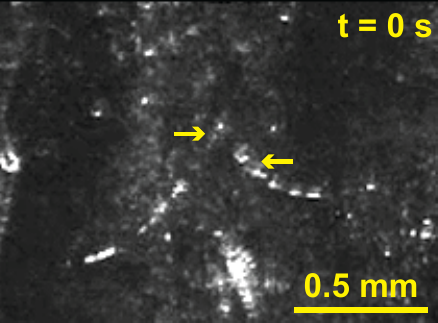}
	\end{subfigure}
	\begin{subfigure}[b]{0.24\textwidth}
		\centering
		\includegraphics*[width=\textwidth]{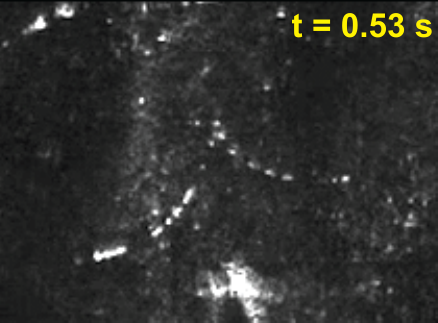}
	\end{subfigure}
    \begin{subfigure}[b]{0.24\textwidth}
		\centering
		\includegraphics*[width=\textwidth]{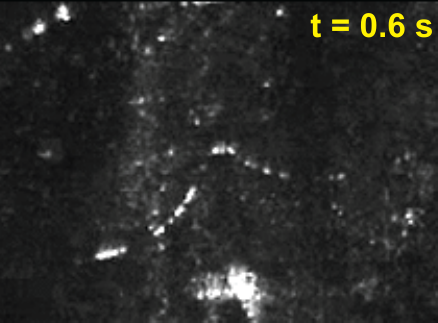}
	\end{subfigure}
    \begin{subfigure}[b]{0.24\textwidth}
		\centering
		\includegraphics*[width=\textwidth]{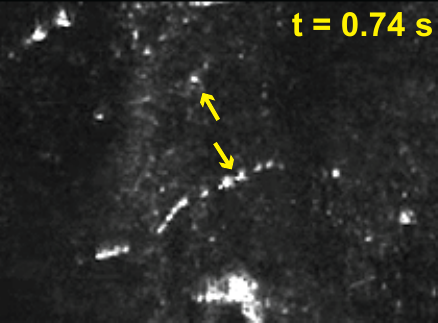}
	\end{subfigure}
	\hfill
    \vspace{0.5cm}
	\begin{subfigure}[b]{0.24\textwidth}
		\centering
		\includegraphics*[width=\textwidth]{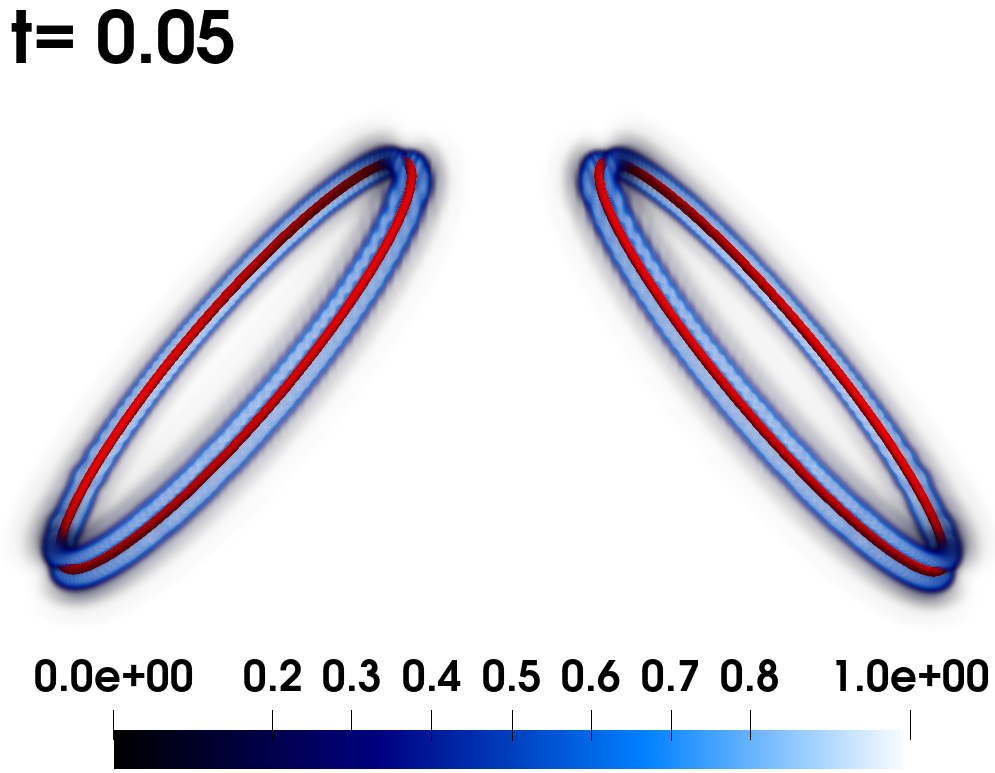}
	\end{subfigure}
	\begin{subfigure}[b]{0.24\textwidth}
		\centering
		\includegraphics*[width=\textwidth]{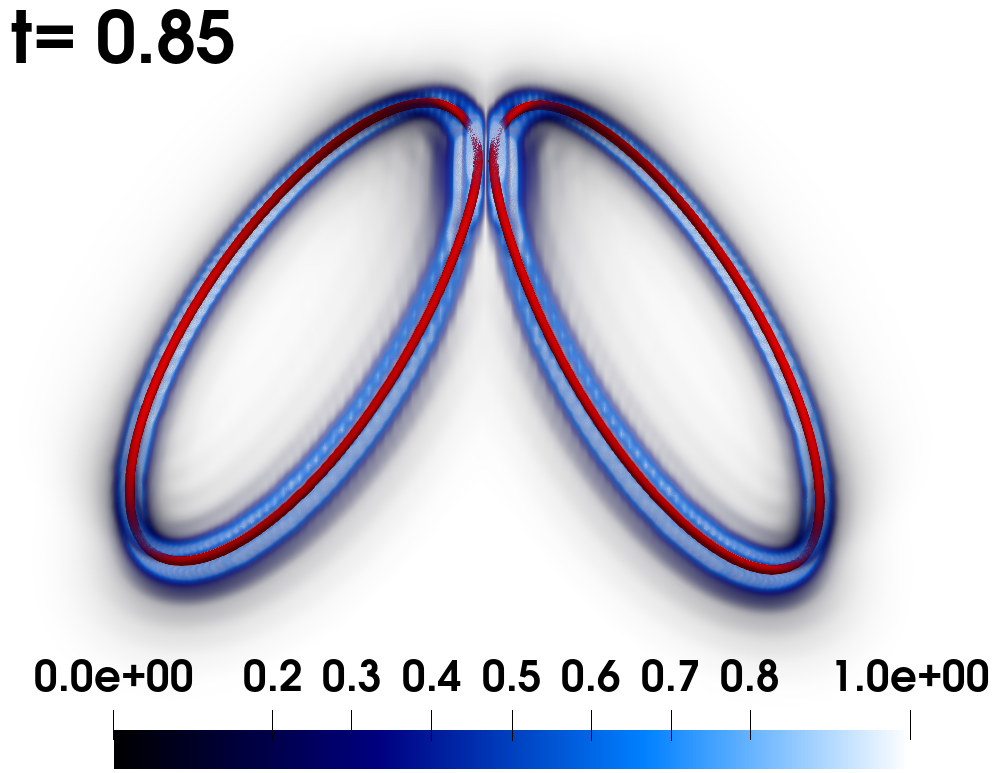}
	\end{subfigure}
    \begin{subfigure}[b]{0.24\textwidth}
		\centering
		\includegraphics*[width=\textwidth]{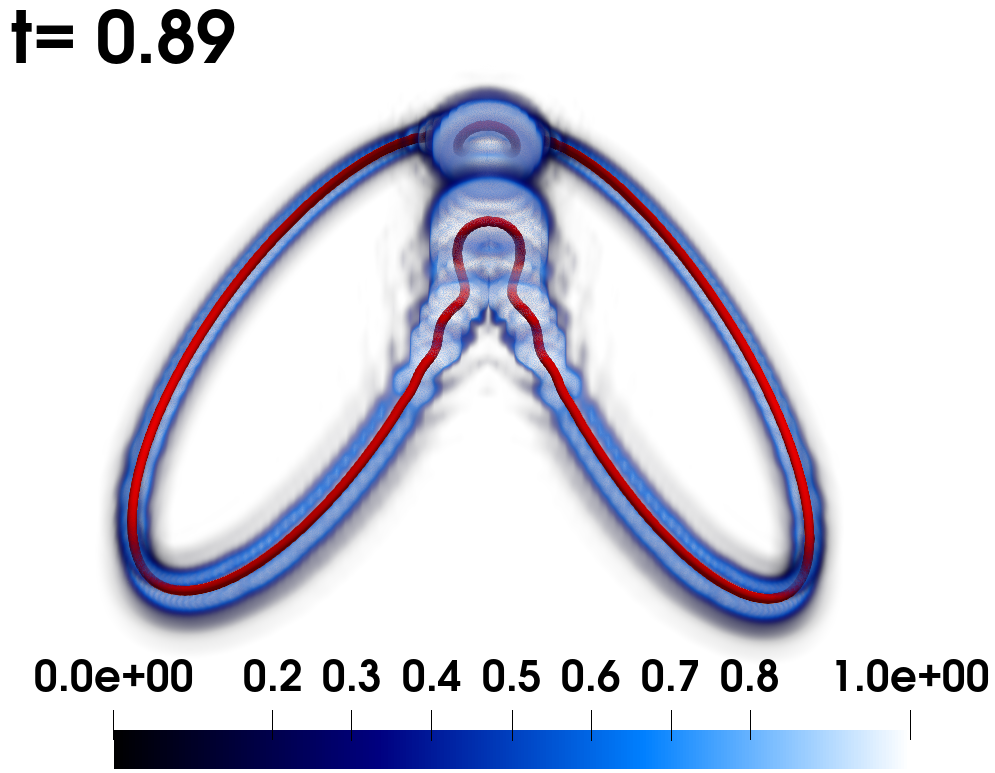}
	\end{subfigure}
    \begin{subfigure}[b]{0.24\textwidth}
		\centering
		\includegraphics*[width=\textwidth]{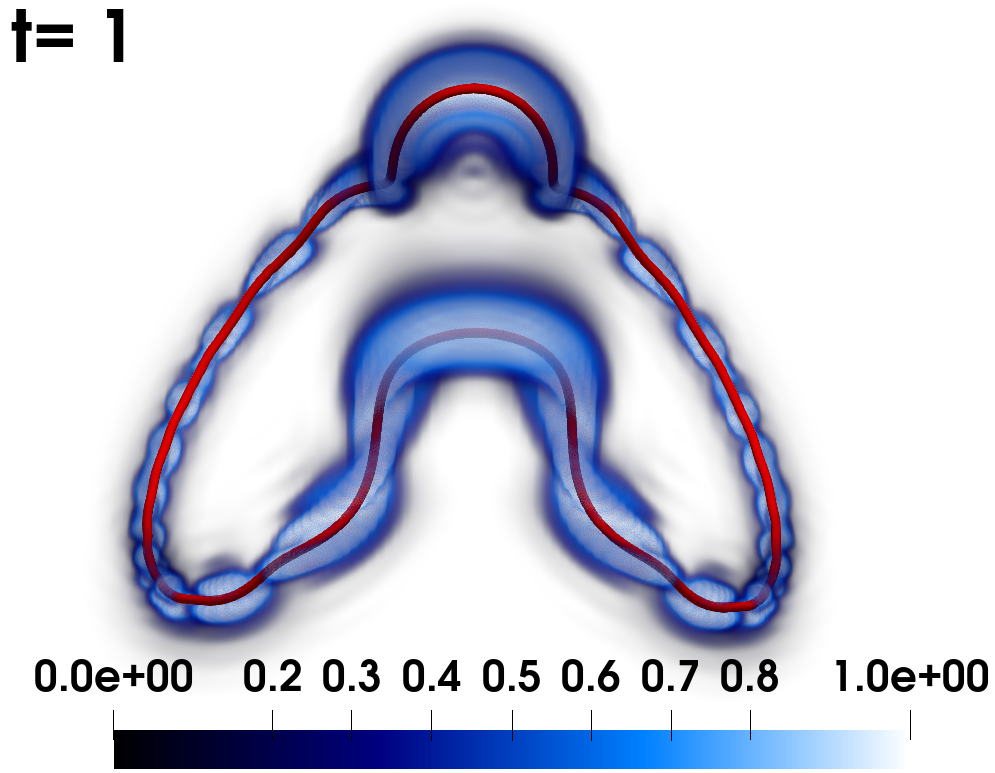}
	\end{subfigure}
\caption{\emph{Top row:} Images showing tracer particles trapped on reconnecting vortices in superfluid helium at 1.65 K. The arrows denote the vortices before and after the reconnection. The first two images show that the vortices approach each other before the reconnection, which occurs at $t=0.58$ s. After the reconnection, the resulting vortices start to move apart, as shown in the last two images.
\emph{Bottom row:} Oblique collision of two circular vortex rings at different 
(dimensionless) times, here in units of $\tau=0.183$s.
The superfluid vortex lines are represented by red tubes (the radius has been greatly 
exaggerated for visual purposes); the scaled normal fluid enstrophy 
$\bom^2/\bom^2_{max}$ is represented by the blue volume rendering.}. 
\label{fig:ring-coll-viz}
\end{figure*}

\paragraph*{Introduction.---} Reconnections are the fundamental events that
change the topology of the field lines in fluids and plasmas during their time evolution. Reconnections thus determine important physical properties, such as mixing and inter-scale energy transfer in fluids \cite{YaoHussainAnnRev2022}, or solar flares and tokamak instabilities in plasmas \cite{Chapman2010}. The nature of reconnections is more clearly studied if the field lines are concentrated in well-separated filamentary structures: vortices in fluids and magnetic flux tubes in plasmas. In superfluid helium this concentration is extreme, providing an ideal context: superfluid vorticity is confined to vortex lines of atomic thickness (approximately $a_0 \approx 10^{-10}~\rm m$); a further simplification 
is that, unlike what happens in ordinary fluids, the
circulation of a superfluid vortex  is constrained to the quantized value $\kappa=h/m=9.97 \times 10^{-8}~\rm m^2/s$, 
where $m$ is the mass of one helium atom and $h$ is Planck's constant. 

It was in this superfluid context that it was theoretically and experimentally recognized
\cite{nazarenko2003,bewley2008,paoletti2010,zuccherQuantumVortexReconnections2012a,villoisUniversalNonuniversalAspects2017a,galantucciCrossoverInteractionDriven2019a,tylutki2021universal}
that reconnections share a universal property irrespective of the initial
condition: the minimum distance between reconnecting 
vortices, $\delta^{\pm}$, scales with time, $t$, according to the form
\begin{equation}
\label{eq:scaling}
	\delta^{\pm}(t) = A^{\pm} (\kappa|t-t_0|)^{1/2},
\end{equation} 
\noindent
where $t_0$ is the reconnection time, and the dimensionless
prefactors $A^-$ and $A^+$ refer respectively to before
($t<t_0$) and after ($t>t_0$) the reconnection. The same scaling law
was then found for reconnections in ordinary viscous fluids 
\cite{yaoSeparationScalingViscous2020}. In the case of a pure
superfluid at temperature $T=0~\rm K$, theoretical work based on
the Gross-Pitaevskii equation (GPE) has shown that
$A^+>A^-$, that is, after the reconnection, vortex lines move away from 
each others faster than in the initial approach; this result has been
related to irreversibility \cite{villoisIrreversibleDynamicsVortex2020,promentMatchingTheoryCharacterize2020}. Indeed, a geometrical constraint imposes 
\cite{promentMatchingTheoryCharacterize2020}
that a piece of vortex length needs to be ``deleted'' 
during the reconnection process. In the GP model, this loss is possible 
by the emission of a rarefaction pulse created immediately after 
the reconnection
\cite{leadbeaterSoundEmissionDue2001b,zuccherQuantumVortexReconnections2012a} which removes some of the kinetic energy and momentum of the vortex configuration.
This vortex energy loss depends on
the ratio $A^+/A^-$, which in turn defines the approaching angle of collision
between the vortices, together with other several geometrical quantities \cite{villoisUniversalNonuniversalAspects2017a,promentMatchingTheoryCharacterize2020}. 
The temporal asymmetry $A^+>A^-$ can be thus interpreted as a non-trivial manifestation of irreversibility, as it originates from an ideal hydrodynamic process independent of the small-scale regularisation mechanism of the fluid.
Indeed, in classical fluid vortex reconnections, although the definition of $A^+$ is more delicate as circulation is not necessarily conserved, the same asymmetry $A^+>A^-$ was reported \cite{yaoSeparationScalingViscous2020}. Instead of the generation of rarefaction pulses, like in the case of $T=0$ superfluids, close to the reconnection, the classical fluid creates a series of thin secondary structures that can be then efficiently dissipated by viscous dissipation.
We note that the reconnection event also generates wavepackets of Kelvin waves about either side of the reconnection cusp, which propagate outward (visible in the bottom panel of Fig~\ref{fig:ring-coll-viz}). These waves are the fundamental mechanism transfering superfluid kinetic energy to smaller scales \cite{vinen2001decay,vinen2002quantum}. The interaction of Kelvin waves with the normal fluid has recently been studied \cite{stasiakCrossComponentEnergyTransfer2024,stasiak2025inverse}.

The aim of this Letter is to investigate the role played by the normal fluid in the reconnection dynamics. In particular, given the temperature dependence of the normal fluid's properties, we study experimentally and numerically the temperature dependence of the prefactors $A^+$ and $A^-$ and numerically investigate the energy injected in the normal fluid. 
To achieve this aim we need a more powerful model than the GPE to account not only for the dynamics of the
superfluid vortices, but also for the dynamics of the normal fluid. 
We show that at non-zero temperatures Eq.~(\ref{eq:scaling}) and the
relation $A^+>A^-$ hold true, in agreement with experiments, revealing, for the 
first time, a temperature dependence of $A^+/A^-$. In addition, we
show that a vortex
reconnection represents an unusual kind of
punctuated energy injection into the normal fluid which acts alongside
the well-known (continual) friction.
When applied to superfluid turbulence, this last result implies that,
if the vortex line density (hence the frequency of reconnections) is large enough, vortex reconnections can maintain the normal fluid in a perturbed state.

\paragraph*{Experimental Method.---} To visualize the reconnection 
dynamics, we decorate the vortices using solidified deuterium 
($\rm{D_2}$) tracer particles of density $202.8~\rm{kg/m^3}$ 
\cite{xu2008density}) and mean radius $1.1\times10^{-6}~\rm m$ 
\cite{tang2023visualization,tang2021visualization}. These particles
are generated by injecting a $\rm D_2$/${}^{4}\rm He$ gas mixture into 
the superfluid helium bath 
\cite{tang2023visualization,fonda2016injection} 
as described in the Supplementary 
Material (SM).
When the particles are near the vortices,
they become trapped inside their cores
because of the Bernoulli pressure arising from the 
circulating superfluid flow.
A thin laser sheet is used to illuminate the particles, and their motion 
is recorded at 200 Hz by a camera positioned at a right angle to the laser 
sheet. A high-quality reconnection event, observed at $T=1.65~\rm K$ 
and capturing both the pre- and post-reconnection dynamics, is shown 
in  Fig.~\ref{fig:ring-coll-viz} as an example. Note that according 
to GP simulations \cite{GiuriatoQuantumVortexReconnections2020} the transfer 
of energy and momentum between particle and vortex does not modify
the approaching rates significantly. Reconnection events reported in this
work have been captured in multiple experiments, either following particle 
injection or long time (\textit{i.e.}, 30-60 s) after towing a grid through superfluid helium. 
(this is also supported by the scaling symmetry 
of the system which allows to draw conclusion for length-scales relevant 
to experiments). 

\begin{figure}[t]
	\centering
	\begin{subfigure}[b]{0.45\textwidth}
		\centering
		\includegraphics*[width=\textwidth]{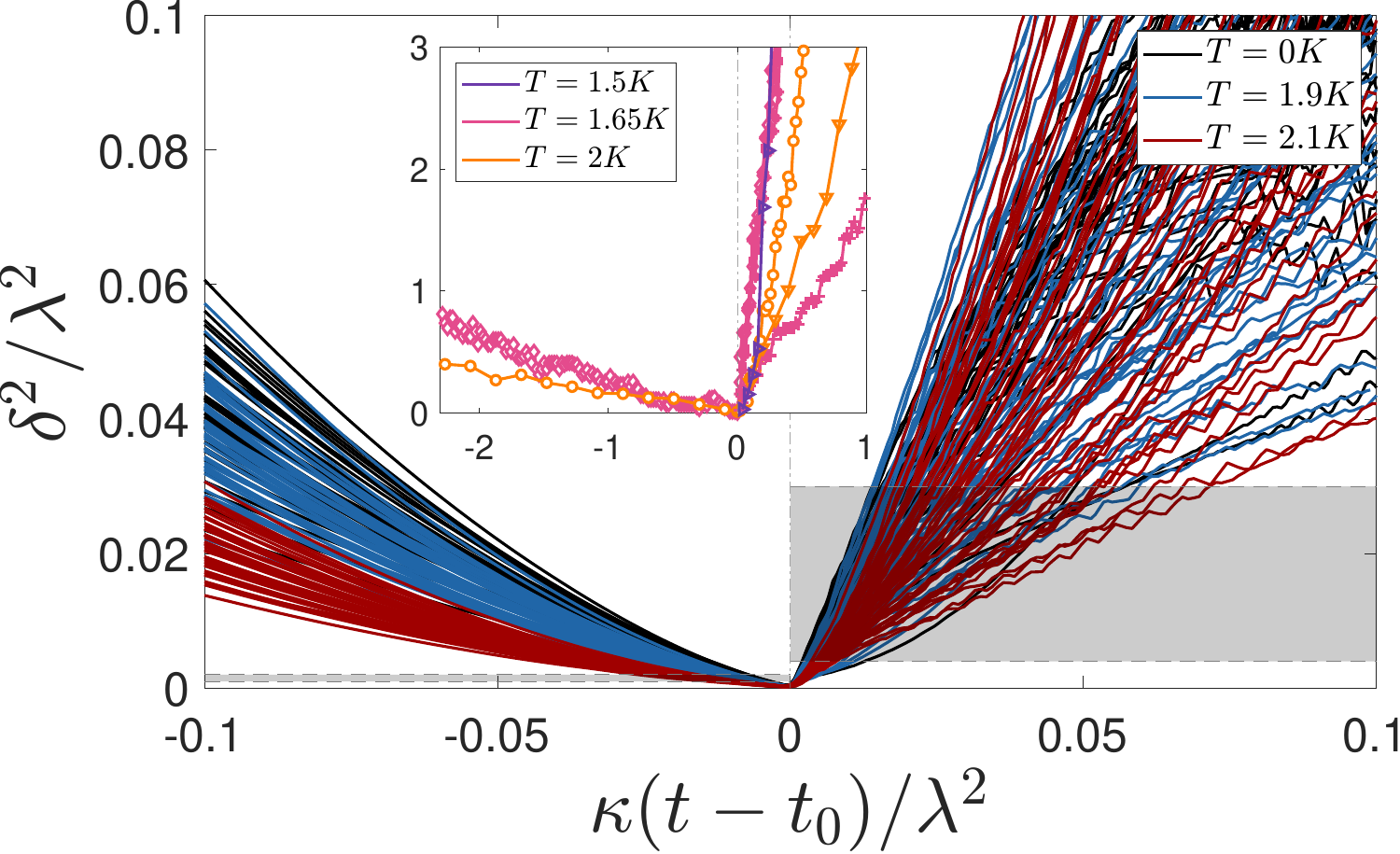}
		\caption{}
		\label{fig:minimum-distance}
	\end{subfigure}
	\begin{subfigure}[b]{0.45\textwidth}
		\centering
		\includegraphics*[width=\textwidth]{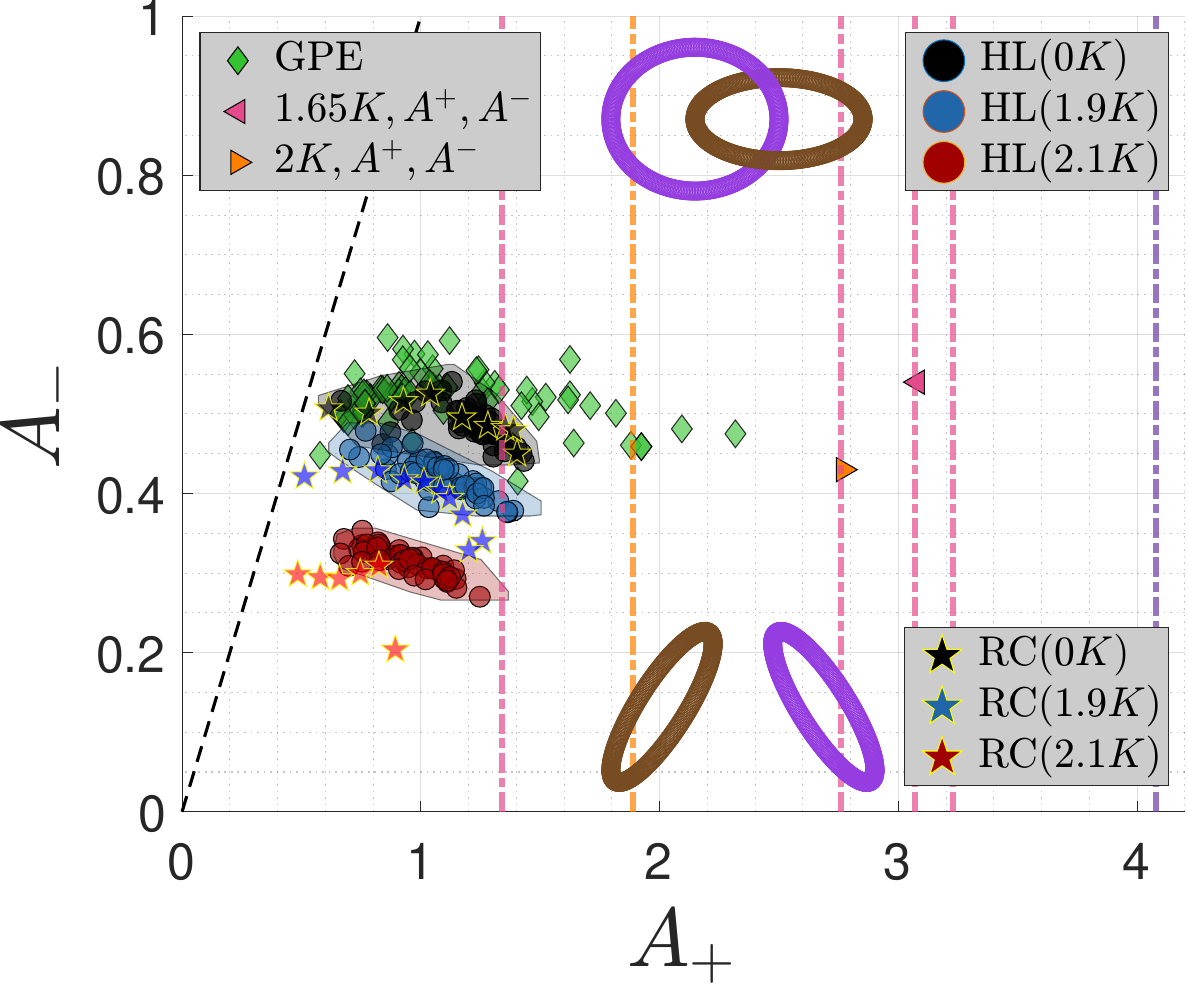}
		\caption{}
		\label{fig:prefactors}
	\end{subfigure}
\caption{\emph{Top row:} Images showing tracer particles trapped on reconnecting vortices in superfluid helium at 1.65 K. The arrows denote the vortices before and after the reconnection. The first two images show that the vortices approach each other before the reconnection, which occurs at $t=0.58$ s. After the reconnection, the resulting vortices start to move apart, as shown in the last two images.
	\emph{Bottom row:} Oblique collision of two circular vortex rings at different 
	(dimensionless) times, here in units of $\tau=0.183$s.
	The superfluid vortex lines are represented by red tubes (the radius has been greatly 
	exaggerated for visual purposes); the scaled normal fluid enstrophy 
	$\bom^2/\bom^2_{max}$ is represented by the blue volume rendering.}
\end{figure}

\paragraph*{Numerical Method.---}We follow the approach of Schwarz \cite{schwarz1988} which exploits
the vast separation of length scales between the vortex core $a_0$ and any
other relevant distance, in particular the average distance between vortices,
$\ell$, in the case of turbulence. Vortex lines are described
as space curves $\s(\xi,t)$ where $\xi$ is arclength. The equation of motion 
of the vortex lines is
\begin{equation}
\label{eq:s}
	\dot{\s}(\xi,t) = \v_s + \frac{\beta}{(1+\beta)}\left[\v_{ns}\cdot \s'\right]\s' + \beta\s'\times\v_{ns}+\beta'\s'\times\left[\s'\times \v_{ns}\right],
\end{equation}
where $\dot{\s}=\partial\s/\partial t$, $\s'=\partial\s/\partial \xi$ 
is the unit tangent vector, 
$\v_n$ and $\v_s$ are the normal fluid and superfluid velocities at $\s$,
$\v_{ns}=\v_n - \v_s$, and $\beta$, $\beta'$ are temperature and Reynolds number dependent 
mutual friction coefficients \cite{galantucciNewSelfconsistentApproach2020b}. The normal fluid velocity $\v_n$ is described as a classical fluid obeying the incompressible ($\nabla\cdot\v_n=0$) Navier-Stokes equations:
\begin{equation}
\label{eq:vn}
	\frac{\partial \v_n}{\partial t} + (\v_n\cdot\nabla)\v_n = 
        -\frac{1}{\rho} \nabla p + \nu_n\nabla^2\v_n + \frac{\mathbf{F}_{ns}}{\rho_n},
\end{equation}
where $\mathbf{F}_{ns}$ is the mutual friction force that couples the normal fluid and the superfluid vortices, and acts as an internal injection mechanism. In Eq.~\eqref{eq:vn}, $\rho=\rho_n + \rho_s$, where $\rho_n$ and $\rho_s$ are the normal fluid and superfluid densities, $p$ is the pressure, 
and $\nu_n$ is the kinematic viscosity of the normal fluid. 
Equations~(\ref{eq:s}) and (\ref{eq:vn})
are solved in dimensionless form by rescaling them by the characteristic time $\tau$ and length $\lambda$. 
The algorithm for vortex reconnections is standard
\cite{baggaleySensitivityVortexFilament2012a}. We consider two distinct initial vortex configurations
at three temperatures $T=0~\rm K$, $1.9~\rm K$ and $2.1~\rm K$
corresponding to the superfluid fractions $\rho_s/\rho=100 \%$,
$58 \%$ and $26 \%$. To compare with experiments, the unit of length is set to $\lambda=1.59 \times 10^{-4}~\rm m$, and the time units to $\tau=0.183~\rm s$ at $T=0~\rm K$ and $1.9~\rm K$, and $\tau=0.242~\rm s$ at $T=2.1~\rm K$, 
see also the SM for details.
All configurations lead to a vortex reconnection.
The first configuration consists of two vortex rings 
of (dimensionless) radius $R\approx 1$ in a tent-like shape
which collide obliquely 
making an initial angle $\alpha$ with the vertical direction, 
as shown in Fig.~\ref{fig:ring-coll-viz}, and,
schematically, in Fig.~\ref{fig:prefactors}.   
By changing the parameter $\alpha$, we create a sample of 12 realizations
at each temperature (again, see the SM for details).
The second configuration is the Hopf link, shown schematically in Fig.~\ref{fig:prefactors}. It consists of two perpendicular linked rings of radius $R\approx1$ with an offset in the $xy$-plane.
By changing the offset, we create a sample of 49 reconnections
at each temperature, as described in the SM.
In all cases, normal fluid structures generated by moving superfluid vortex rings \cite{kivotides-barenghi-samuels-2000}, are initially prepared to eliminate any potential transients.

\paragraph*{Scaling law. ---}In the experiment, two reconnections were observed where both $A^+$ and $A^-$ could be identified and calculated, at $T=1.65K$ and $T=2K$, plotted as orange triangles in Fig.~\ref{fig:prefactors}. We also analysed six additional experimental observations of the \emph{post-reconnection} dynamics only (vertical dot-dashed lines). All were consistent with the $\delta^{1/2}$ scaling with $A^+$ in the range 1.2-4.2, plotted as vertical lines in Fig.~\ref{fig:prefactors} Their corresponding minimal distances are displayed in the inset of Fig~\ref{fig:minimum-distance}.
The pre-reconnection factor $A^-$ lies within the 0.4-0.6 range, consistent with the results of the numerics, and a clear temperature effect between a superfluid component majority and normal fluid component majority.
In the case of the Hopf link 
we have performed 147 simulations 
(49 across 3 temperatures) as shown in Fig.~\ref{fig:minimum-distance} and
verified Eq.~(\ref{eq:scaling}) for the minimum distance $\delta^{\pm}$. 
The prefactors $A^{\pm}$ have been computed in the shaded region 
of the figure. In the pre-reconnection regime ($t<t_0$) we observe
a clear segregation of the values of $A^-$ due to temperature: 
the minimum distance grows more rapidly with time if the temperature is
lowered. 
In stark contrast, there is almost no memory of the temperature in the
post-reconnection regime ($t>t_0$). 

At $T=0~{\rm K}$, our calculations for superfluid helium (black symbols in Fig.~\ref{fig:prefactors}) are in good agreement with previous results obtained with the GPE \cite{villoisIrreversibleDynamicsVortex2020} (green diamonds), showing irreversible dynamics. In addition, the computed values of $A^-\approx 0.4$-$0.6$ at $T=0~{\rm K}$ are consistent with analytical calculations \cite{boue-etal-2013,rica-2019}. At non-zero temperatures, our results confirm the irreversibility of vortex reconnections observed at $T=0$ as $A^+$ is always larger than $A^-$. Importantly, this asymmetry is recovered in all our simulations, regardless of their initial condition.
The same asymmetry between $A^+$ and $A^-$ at non-zero temperatures has been observed for reconnections in finite-temperature
Bose-Einstein condensatates \cite{allen2014}, although in this work the system is not homogeneous (the condensate is confined by a harmonic trap) and the thermal component is a ballistc gas, not a viscous fluid. Note that the vortex reconnections in classical viscous fluids reported in 
\cite{yaoSeparationScalingViscous2020} also display
a clear $1/2$ power-law scaling for the minimum distance
with $A^- \approx 0.3$-$0.4$, which again shows good agreement with our results. The scaling law (Eq.~\ref{eq:scaling}) and the range of values of $A^-$ hence appear to have a universal character in vortex reconnections, independently of the nature of the fluid, classical or quantum.

\begin{figure}[h]
	\centering
	\includegraphics*[width=0.45\textwidth]{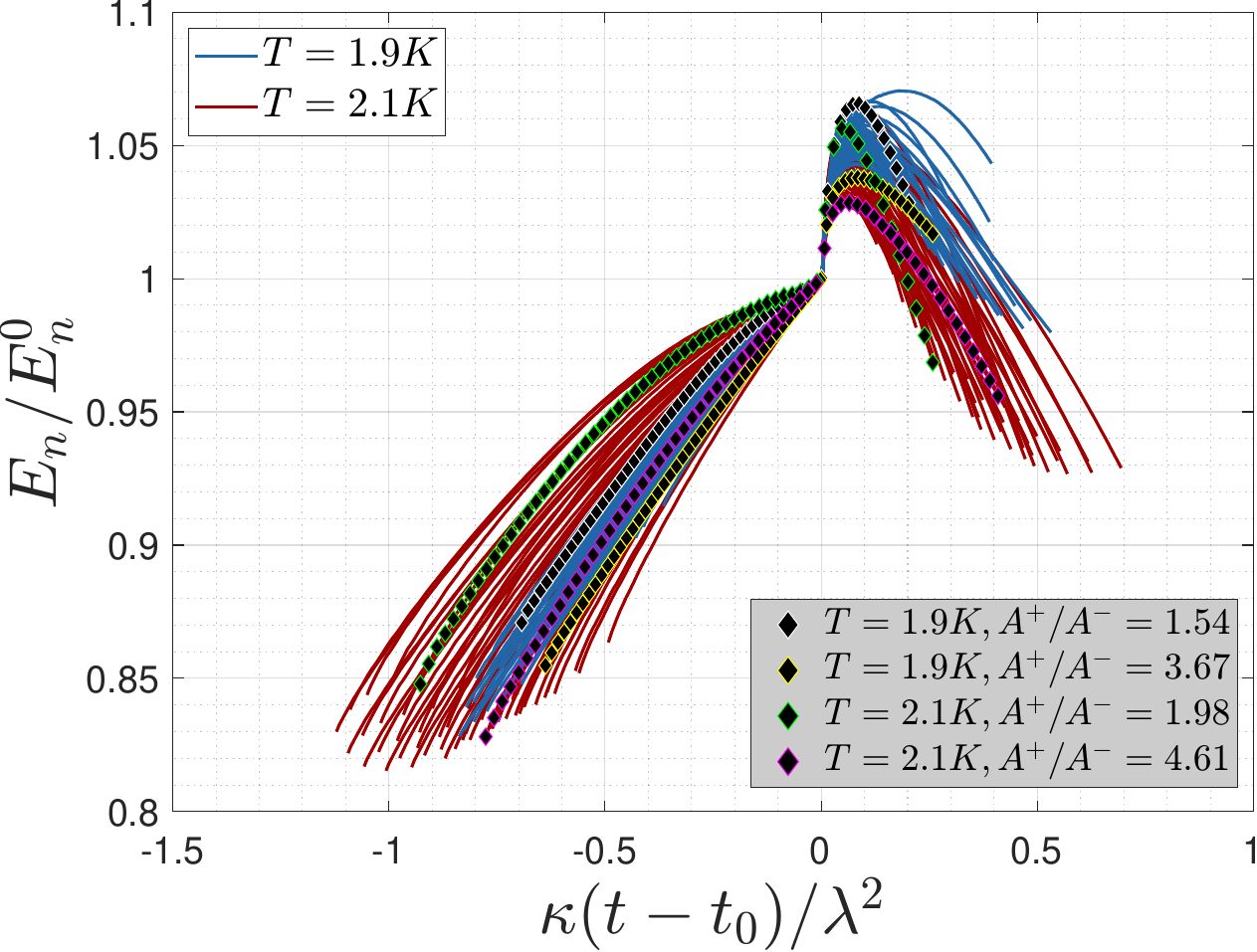}
	\caption{Normal fluid kinetic energy $E_n$ scaled by $E_n^0$ (the 
kinetic energy at $t=t_0$), plotted versus (dimensionless)
$\kappa (t-t_0)$ for the Hopf link reconnections. Black diamonds represent the simulations 
with minimum and maximum prefactor ratios $A^+/A^-$ at $T=1.9~\rm K$ and 
$T=2.1~\rm K$ respectively.}
	\label{fig:energy-evol}
\end{figure}

\paragraph*{Energy injection. ---}
The normal fluid impacts the dynamics of reconnecting superfluid vortices via the temperature dependent mutual friction coefficients. Conversely, the motion of superfluid vortices involved in the reconnection process influence, significantly, the dynamics of the normal fluid. Figure~\ref{fig:energy-evol} indeed shows that the normal fluid energy, $E_n$,
suddenly increases at the reconnection time by an amount ($\approx 5\%$)
which is smaller but comparable to the continuous energy increase as vortex
lines approach each other. Indeed the
curvature $\zeta=|\s''|$ of the vortex line 
spikes at $t=t_0$ when
the reconnection cusp is created, and, in the first approximation \cite{galantucci-krstulovic-etal-2023},
the magnitude of the energy injected in the normal fluid per unit time $I$
is proportional to the strength of the mutual friction force $\mathbf{F}_{ns}$ which scales as
$|\mathbf{F}_{ns}(\s)|\propto|\dot{\s}-\v_n|\propto|\dot{\s}|\propto\zeta$. This sudden
transfer of energy
\cite{stasiakCrossComponentEnergyTransfer2024} from the superfluid
vortex configuration to the normal fluid is the origin of the 
small scale normal fluid enstrophy structures
which are visible in Fig.~\ref{fig:ring-coll-viz}.


\begin{figure}
	\centering
	\includegraphics*[width=0.48\textwidth]{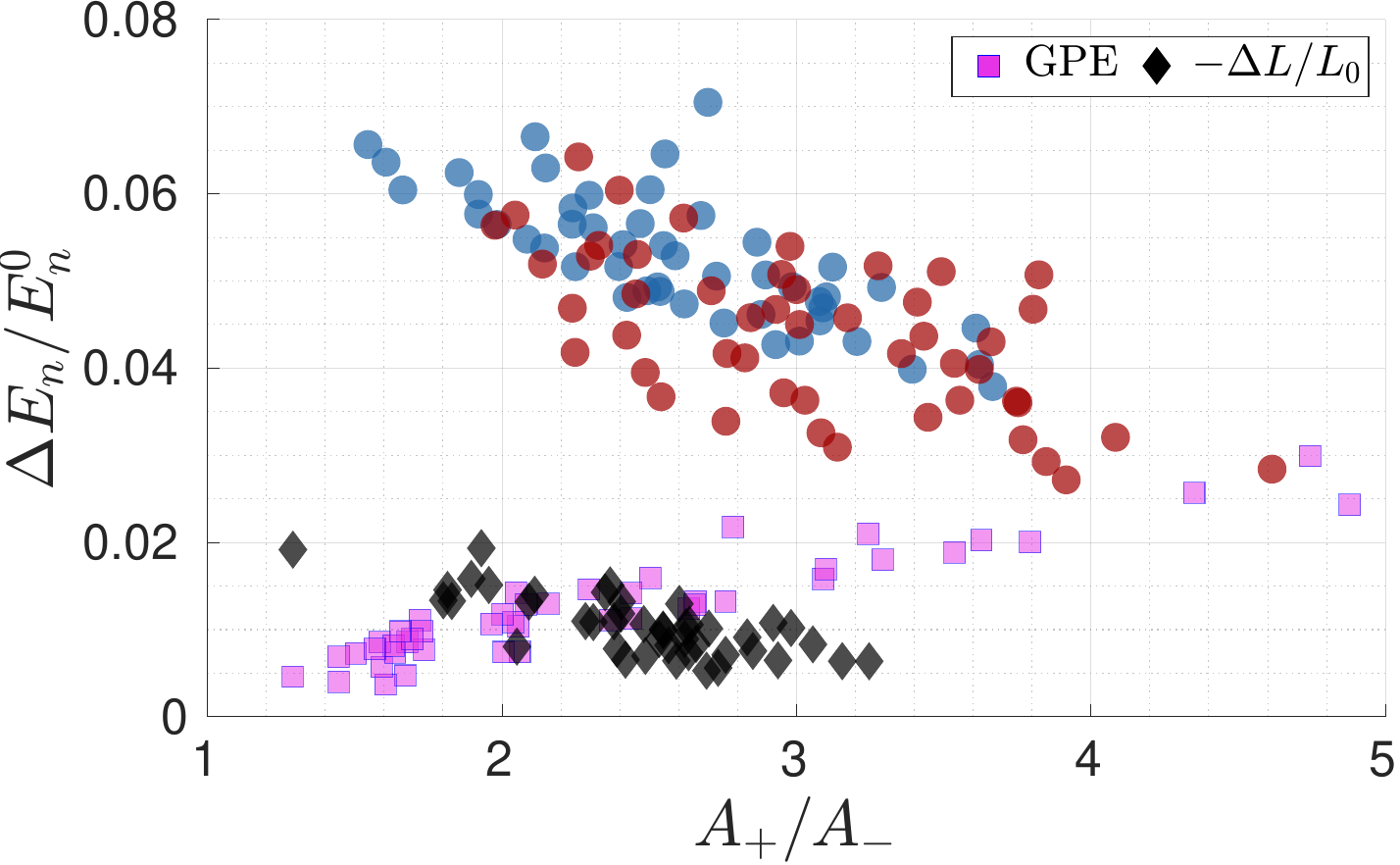}
	\caption{Normalized energy jumps $\Delta E_n/E_n^0$ for Hopf link 
reconnections.The solid black diamonds are the normalized change in line length $\Delta L/L_0$ in the
$T=0~\rm K$ case. Blue and red circle correspond to $T=1.9K$ and $T=2.1K$ respectively. The purple squares are from GPE simulations of
Villois et al. \cite{villoisIrreversibleDynamicsVortex2020}.}
	\label{fig:energy-jumps}
\end{figure}

TThe total energy injected into the normal fluid by the reconnection,
$\Delta E_n$, which hereafter
we refer to as the energy jump, is defined as
\begin{equation}
	\Delta E_n = \max{\left[E_n(t>t_0)\right]} - E_n^0,
\end{equation} 
where $E_n^0=E_n(t_0)$ is the normal fluid kinetic energy at $t=t_0$.
Normalized energy jumps are plotted in
Fig.~\ref{fig:energy-jumps} as a function of
the ratio $A^+/A^-$. Here, we observe that the larger $A^+/A^-$ is,
the smaller the normal fluid excitation is.

The emission of the sound pulse at the vortex reconnection 
\cite{leadbeaterSoundEmissionDue2001b} which is typical of the GPE model
is absent in our incompressible hydrodynamic approach. To model
this effect, the change of vortex length, $\Delta L$, created by the
vortex reconnection algorithm is always negative by construction
\cite{baggaleySensitivityVortexFilament2012a},
because, in the local induction approximation to the Biot-Savart law, 
the superfluid incompressible kinetic energy, $E_s$, is proportional to the vortex length,
$L$. Such procedure ensures that at $T=0~{\rm K}$ when a reconnection occurs
$\Delta E_s\propto \Delta L < 0$.
Consequentially, in the absence of any dissipative normal fluid, 
the superfluid energy 
$E_s$ that would be transferred to the sound pulse, normalized with its value $E_s^0$ at 
reconnection, is $-\Delta L/L_0$. 
If these normalized energy jumps (black diamonds
in Fig.~\ref{fig:energy-jumps}) are compared to the results obtained with the 
compressible GPE \cite{villoisIrreversibleDynamicsVortex2020} (purple squares)
we find a good agreement, confirming that the model we employ, 
is suitable for the investigation of the features of single reconnection events.

\paragraph*{Implications for turbulence. ---}
Our numerical results have implications for our understanding of 
quantum turbulence \cite{BSS2023}.
A fully developed turbulent tangle of vortices is
characterized by its vortex line density $\mathcal{L}$ (vortex length
per unit volume); the frequency 
of vortex reconnections per unit volume is 
${f=(\kappa/6\pi)\mathcal{L}^{5/2}\ln(\mathcal{L}^{-1/2}/a_0)}$
\cite{barenghi2004}. From Fig.~\ref{fig:energy-evol} we estimate the 
normal fluid reconnection relaxation time $\tau_n$ as the time 
after reconnection at which the normal fluid energy $E_n/E_0$ has decayed
to the pre-reconnection level: in our dimensionless units, $\kappa \tau_n \approx 0.25$. 
Using this timescale, we estimate that
the average vortex line density that is required to sustain the normal fluid 
in a perturbed state via frequent vortex reconnections is approximately
$\mathcal{L} \approx 10^7$ to $10^8\mathrm{m}^{-2}$. 
Experiments in $^4$He
\cite{schwarz1981,milliken1982,roche2008,roche2007,Babuin2014} and in $^3$He
\cite{bradley2006} can achieve vortex line densities much larger than this.

Above the vortex line density threshold, the increase of normal fluid energy generated by the reconnection will not have time to decay before the subsequent reconnection occurs, which will add again more energy. In this manner, the normal fluid energy will not decay to zero, but will increase in time. In general, such finite amplitude normal fluid disturbances constantly injected by reconnections may become relevant for various superfluid helium systems. One example is the oscillatory flows, which are widely studied in superfluid helium using vibrating wires and forks. At low frequency, perturbations which start at finite-amplitude rather than infinitesimal-amplitude level have enough time to become of order one (hence visible and destabilizing) in the supercritical part of the cycle \cite{barenghi1989modulated}. A second example is pipe flow, again relevant to helium experiments, which is known to suffer from finite-amplitude instabilities \cite{peixinho2007finite}. This effect clearly needs further investigations.

\paragraph*{Conclusions.---} We have conducted an experiment using passive particle tracers and a suite of numerical simulations of vortex reconnections over a wide range of temperatures using a model of $^4$He which accounts for the coupled dynamics of superfluid and normal fluid components.
We have verified the scaling law of the minimum vortex distance 
$\delta^{\pm}=A^{\pm} (\kappa |t-t_0|)^{1/2}$ and found that the approach prefactor $A^-$ has a clear temperature dependence independent of the geometry in both experiments and numerics, in contrast to the
separation prefactor $A^+$. The prefactors are in good agreement
with GPE simulations \cite{villoisIrreversibleDynamicsVortex2020,allen2014} 
and classical fluid reconnections \cite{yaoSeparationScalingViscous2020}
revealing that vortex reconnections display a universal behaviour, linked to irreversible vortex energy dissipation, regardless of the nature
of the fluid (classical or quantum) and of temperature, \textit{i.e.} regardless of the small scale energy transfer mechanism. 
It is worth noting that the behaviour, as a function of $A^+/A^-$, of 
the energy injected in the normal fluid (at $T>0$) and of the energy transferred to sound (at $T=0$)
\cite{villoisIrreversibleDynamicsVortex2020,leadbeaterSoundEmissionDue2001b} is dissimilar: the former decreases as $A^+/A^-$
increases, the latter the opposite. This likely arises from the distinct physics governing the loss of superfluid 
incompressible kinetic energy: mutual friction at $T>0$, quantum pressure at $T=0$.
We have also found that a reconnection event suddenly injects an amount of energy 
into the normal fluid which is comparable to the energy transferred by friction
during the vortex approach. Applying these results to turbulence, we have
compared the decay time of the normal fluid structures created by a
reconnection to the frequency of reconnections in a vortex tangle, and argued
that, if the vortex line density is large enough, these punctuated
energy injections should sustain the normal fluid in a perturbed state, which may lead to a new type of turbulence.

\paragraph*{Acknowledgements.---} Y.M.X., Y.A., and W.G. acknowledge the support from the Gordon and Betty Moore Foundation through Grant DOI 10.37807/gbmf11567 and the US Department of Energy under Grant DE-SC0020113. The experimental work was conducted at the National High Magnetic Field Laboratory at Florida State University, which is supported by the National Science Foundation Cooperative Agreement No. DMR-2128556 and the state of Florida. G.K. acknowledges financial support from the Agence Nationale de la Recherche through the project QuantumVIW ANR-23-CE30-0024-02. P.Z.S. acknowledges the financial support of the UCA ``visiting doctoral student program'' on complex systems. Computations were carried out at the Mésocentre SIGAMM hosted at the Observatoire de la Côte d’Azur.

\appendix
\section*{Supplementary Materials}

\begin{figure*}[t]
	\centering
	\includegraphics[width=0.8\textwidth]{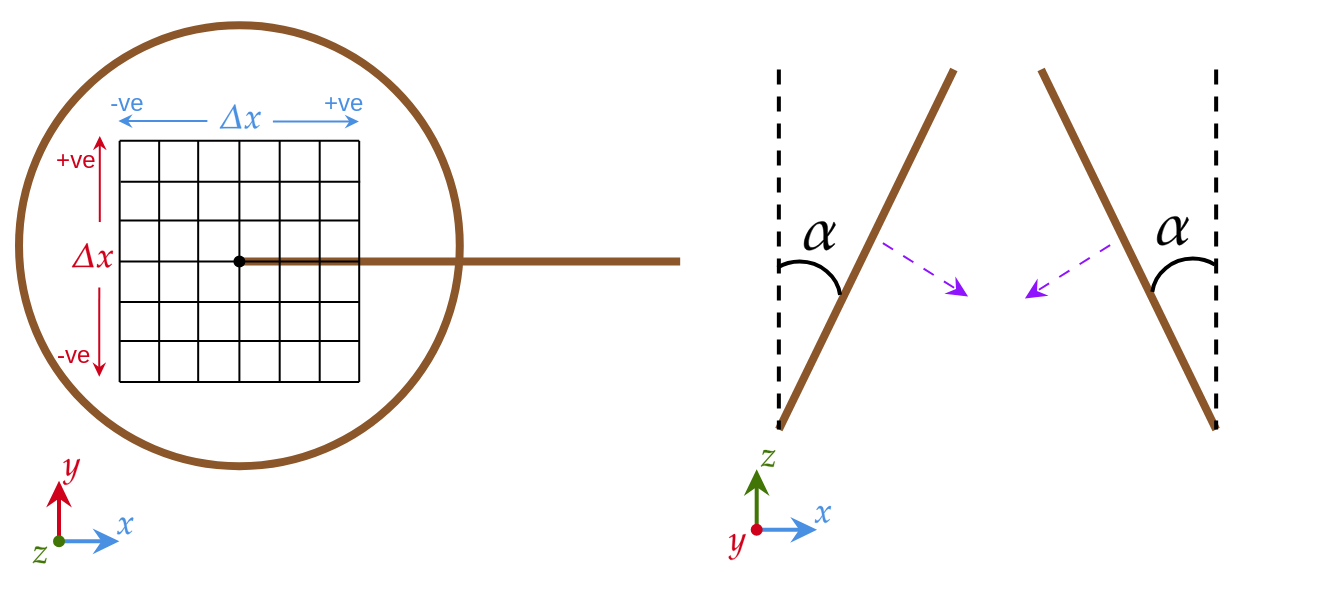}
	\caption{Schematic diagram for numeric initial condition. \emph{Left:} Hopf-link. \emph{Right:} Oblique collision.}
	\label{fig: ring-coll-viz}
\end{figure*}

\section{Experimental Method}

To produce solidified deuterium ($\rm D_2$) tracer particles, we slowly inject a gas mixture of $5\%$ $\rm D_2$ and $95\%$ ${}^{4}\rm He$ into a superfluid helium bath. Our gas injection system is similar to that described by Fonda et al. \cite{fonda2016injection}. A solenoid valve is installed to control the duration of the gas injection, and a needle valve is used to regulate the gas flow rate. The injected $\rm D_2$ gas solidifies into small ice particles with a mean radius of 1.1 $\mu \rm m$, derived from the particle settling velocities in quiescent superfluid helium \cite{tang2023visualization}. A 473 nm continuous-wave laser sheet (thickness: $0.8~\rm mm$) illuminates the particles, and their motion in the laser sheet plane is recorded by a camera at 200 frames per second with a maximum resolution of $2560 \times 1440$ pixels. We then identify vortex reconnection events from the recorded videos, and manually track the coordinates of the trapped particles for the pre- (if captured) and post-reconnection. Knowing the particle coordinates, the minimum distance between the reconnecting vortices $\delta^{\pm}(t)$ can be measured. We calculate the prefactors $A^\pm$ using the slopes of linear fits to the $\delta^2(t)$ data.

\section{Numerical Method}

Using Schwarz mesoscopic model \cite{schwarz1988}, vortex lines can be described as space curves $\s(\xi,t)$ of infinitesimal thickness, with a single quantum of circulation $\kappa=h/m_4=9.97\times10^{-8}\text{m}^2/\text{s}$, where $h$ is Planck's constant, $m_4=6.65\times10^{-27}\text{kg}$ is the mass of one helium atom, $\xi$ is the natural parameterisation, arclength, and $t$ is time. These conditions are a good approximation, since the vortex core radius of superfluid \textsuperscript{4}He($a_0=10^{-10}\text{m}$) is much smaller than any of the length scale of interest in turbulent flows. The equation of motion is
\begin{equation}
	\dot{\s}(\xi,t) = \v_s + \frac{\beta}{1+\beta}\left[\v_{ns}\cdot \s'\right]\s' + \beta\s'\times\v_{ns}+\beta'\s'\times\left[\s'\times \v_{ns}\right],
\end{equation}
where $\dot{\s}=\partial\s/\partial t$, $\s'=\partial\s/\partial \xi$ is the unit tangent vector, $\v_{ns}=\v_n - \v_s$, $\v_n$ and $\v_s$ are the normal fluid and superfluid velocities at $\s$ and $\beta$,$\beta'$ are temperature and Reynolds number dependent mutual fricition coefficients \cite{galantucciNewSelfconsistentApproach2020b}. The superfluid velocity $\v_s$ at a point $\x$ is determined by the Biot-Savart law
\begin{equation}
	\v_s(\x,t) = \frac{\kappa}{4\pi}\oint_{\mathcal{T}}\frac{\s'(\xi,t)\times\left[\x-\s(\xi,t)\right]}{|\x-\s(\xi,t)|}d\xi,
\end{equation}
where $\mathcal{T}$ represents the entire vortex configuration.
There is currently a lack of a well-defined theory of vortex reconnections in superfluid helium, like for the Gross-Pitaevskii equation \cite{villoisIrreversibleDynamicsVortex2020,villoisUniversalNonuniversalAspects2017a,promentMatchingTheoryCharacterize2020}. An \emph{ad hoc} vortex reconnection algorithm is employed to resolve the collisions of vortex lines \cite{baggaleySensitivityVortexFilament2012a}.

A \emph{two-way model} is crucial to understand the accurately interept the back-reaction effect of the normal fluid on the vortex line and vice-versa \cite{stasiakCrossComponentEnergyTransfer2024}. We self-consistently evolve the normal fluid $\v_n$ with a modified Navier-Stokes equation
\begin{equation}
	\frac{\partial \v_n}{\partial t} + (\v_n\cdot\nabla)\v_n = -\nabla\frac{p}{\rho} + \nu_n\nabla^2\v_n + \frac{\mathbf{F}_{ns}}{\rho_n},
\end{equation}
\begin{equation}
	\mathbf{F}_{ns} = \oint_{\mathcal{T}}\mathbf{f}_{ns}\delta(\x-\x)d\xi, \quad \nabla\cdot\v_n=0,
\end{equation}
where $\rho=\rho_n + \rho_s$ is the total density, $\rho_n$ and $\rho_s$ are the normal fluid and superfluid densities, $p$ is the pressure, $\nu_n$ is the kinematic viscosity of the normal fluid and $\mathbf{f}_{ns}$ is the local friction per unit length \cite{galantucciCoupledNormalFluid2015a}
\begin{equation}
	\mathbf{f}_{ns} = -D\s'\times\left[\s'\times(\dot{\s}-\v_n)\right]-\rho_n\kappa\s'\times(\v_n-\dot{\s}), 
\end{equation}
where $D$ is a coefficient dependent on the vortex Reynolds number and intrinsic properties of the normal fluid.

The results in this Letter are reported in dimensionless units, where the characteristic length scale is $\tilde{\lambda}=D/D_0$, where $D^3=(1\times10^{-3}\text{m})^3$ is the dimensional cube size, $D_0^3=(2\pi)^3$ is the non-dimensional cubic computational domain. The time scale is given by $\tilde{\tau} = \tilde{\lambda}^2\nu_n^0/\nu_n$, where the non-dimensional viscosity $\nu_n^0$ resolves the small scales of the normal fluid. In these simulations, these quantities are $\tilde{\lambda}=1.59\times10^{-4}\text{cm}$, $\nu_n^0=0.16$ and $\tilde{\tau} = 0.183$s at $T=0K$ and at $T=1.9K$ and $\tilde{\tau}=0.242$s at $T=2.1K$.
We consider two distinct initial vortex geometries at $T=0K,1.9K$ and $2.1K$. The first is a Hopf link, two linked rings of radius $R\approx1$ with an offset in the $xy$-plane defined by parameters $\Delta l_x$ and $\Delta l_y$. The offsets are chosen so that $(\Delta l_x, \Delta l _y) \in \lbrace(0.125i,0.125j)|i,j=-3,\cdots,3 \rbrace$, a total of 49 reconnections for each temperature. The second geometry is a collision of vortex rings of radius $R\approx1$ in a tent-like configuration (see Fig. \ref{fig: ring-coll-viz}), making an angle $\alpha$ with the vertical. We take 12 realisations of $\alpha$, such that $\alpha\in\lbrace i\pi/13|i=1,\cdots,12\rbrace$. 

In both cases, normal fluid rings are initially superimposed to match the vortex lines, eliminating the transient phase of generating normal fluid structures. The Lagrangian discretisation of vortex lines is $\Delta\xi=0.025$ (a total of 668 discretiation points) with a timestep of $\Delta t_{VF}=1.25\times10^{-5}$. A total of $N=256^3$ Eulerian mesh points were used for the normal fluid, with a timestep of $\Delta t_{NS} = 40\Delta t_{VF}$.


\end{document}